# Quenching of superconductivity by Co doping in $K_{0.8}Fe_2Se_2$

Tingting Zhou, Xiaolong Chen*, Jiangang Guo, Gang Wang, Xiaofang Lai, Shunchong Wang, Shifeng Jin, and Kaixing Zhu

Research & Development Center for Functional Crystals, Beijing National Laboratory for Condensed Matter Physics, Institute of Physics, Chinese Academy of Sciences, P.O. Box 603, Beijing 100190, China

*E-mail: chenx29@aphy.iphy.ac.cn.

We synthesized a series of $K_{0.8}Fe_{2-x}Co_xSe_2$ samples with nominal compositions $0 \leq x \leq 0.035$ and investigated their physical properties. The results show that the superconductivity in $K_{0.8}Fe_{2-x}Co_xSe_2$ is quenched down to 5 K by 0.5 at. % Co doping, the fastest quenching rate ever-reported. The role played here by Co is in contrast with the one in FeAs based superconductors where Co usually induces superconductivity from parent compounds. Such a rapid quenching favors a localized 3$d$ model against the itinerant one for iron pnictide superconductors.

Similar to the cuprates, superconductivity emerges in iron-based superconductors when AFM order is suppressed by carrier doping or by application of pressure. In this context, the family of the iron-based superconductors has been expanded rapidly by discovering LaOFeAs (F doped) (1111 phase) [1], $AFe_2As_2$ (A=K, Sr, Ba) (122 phase) [2-4], LiFeAs (111 phase) [5], and FeSe (11 phase) [6]. Among them, the anti-PbO-type FeSe was paid special attention for its substantially simplified structure, meanwhile without arsenic and less toxic. The partial Te substitution for Se enhances the superconducting transition temperature ($T_c$) up to 15.2 K [7]. Simultaneously, $T_c$ of FeSe could be raised up to 37 K by the exertion of high pressure [8]. These results imply that the superconductivity of FeSe-based compounds is sensitive to the carriers and structure modulation. Moreover, the angle resolved photoemission spectroscopy (ARPES) showed that the normal state of $FeSe_{0.42}Te_{0.58}$ is a strongly correlated metal, which is significantly different with the 1111 and 122 iron pnictide families [9]. Very recently, $K_{0.8}Fe_2Se_2$ was found to become superconductive at around 30K at ambient pressure [10]. Soon afterwards, $A_{0.8}Fe_2Se_2$ (A= Cs, Rb) have been synthesized by replacing K with other alkali metals (Cs, Rb) with similar transition temperatures [11,12]. Subsequent works showed that the superconductivity in this system is closely related to the orderings of Fe vacancies and the anomalous "hump" in the resistivity [13,14]. Large moment of 3.31$\mu_B$/Fe antiferromagnetic (AFM) order with Neel temperature as high as 559 K and high $T_c$ superconductivity coexist in $K_{0.8}Fe_{1.6}Se_2$ [15]. The newly discovered Fe-based superconductors, $A_{0.8}Fe_2Se_2$, are attracting immense interests owing to its peculiar properties.

Doping on the Fe sites is thought to be an efficient way to get a deep insight into the superconducting mechanism [16]. Among the substitution alternatives, transition metals with unpaired $3d$ electrons and with comparable ionic radii of Fe were paid much attention to. For example, cobalt has been widely used as a dopant, which adds carriers directly into the FeAs layer, to induce superconductivity in 1111 and 122 type iron pnictides, such as LaFeAsO ($T_c$=14.3 K) [17], CaFeAsF ($T_c$=22 K) [18], $SrFe_2As_2$

($T_c$=20 K) [16], and BaFe$_2$As$_2$ ($T_c$=22 K) [19]. In these systems, superconductivity seemed to have considerable tolerance of disorder in FeAs layers for the high Co content (10 at. %- 20 at. %). However, for FeSe, it was noticed that superconductivity was destroyed completely at a relatively low Co substitution (5 at. %) [20,21]. Such a strong suppression of the superconductivity was considered to be correlated to the electron doping which modifies the Fermi surface, resulting in the collapse of the nesting [22].

In this Letter, we systematically investigated the structure and transport properties of a series of Co-substituted K$_{0.8}$Fe$_2$Se$_2$ compounds. It was found that superconductivity was rapidly quenched down to 5 K with a very dilute substitution of Co, ~0.5 at. %, the fastest one in iron-based superconductors, reminiscent of the behavior observed in the cuprates. The possible mechanism for the suppression of superconductivity was studied and a localized 3$d$ model may be responsible for it.

A series of crystal-like K$_{0.8}$Fe$_{2-x}$Co$_x$Se$_2$ samples(nominal compositions x=0, 0.005, 0.015, 0.02, 0.025, 0.03, 0.035) were prepared by a two-step solid-state reaction method. First, Fe$_{1-x/2}$Co$_{x/2}$Se compounds were synthesized using high-purity powders of iron (Alfa, 99.9+%), cobalt (Alfa, 99.5%), and selenium (Alfa, 99.99%) by a similar method that was described in Ref. 6. Then, Fe$_{1-x/2}$Co$_{x/2}$Se and K (Sinopharm Chemical, 97%) mixed with predetermined stoichiometry were put into alumina crucibles and sealed into silica ampoules under an atmosphere of 0.2 bar argon. The mixtures were heated to 1040℃ and maintained for 2 hours, then cooled down to 800℃ with a rate of 4 ℃/h and finally cooled down to room temperature with a rate of 100 ℃/h. The morphologies and phase compositions of the samples were observed by scanning electron microscopy (SEM, XL30 S-FEG) and induced coupling plasma-atomic emission spectrum (ICP-AES). Phase identifications and structures of the materials were examined by powder x-ray diffraction (PXRD) at room temperature using a panalytical X'pert diffractometer with Cu $K\alpha$ radiation. For PXRD measurement, pieces of crystals were cleaved and powdered. Rietveld refinements were performed by using the FULLPROF package [23]. The electrical resistivity measurements were performed by a standard four-probe method on the physical property measurement system (PPMS, Quantum Design). Platinum wires were attached onto the fresh surface with silver paste. The dc magnetic susceptibilities were measured by vibrating sample magnetometer (VSM, Quantum Design).

The elemental compositions of the as-prepared samples were measured by ICP-AES and were listed in Table Ⅰ. The measured Co content is lower than the nominal one. The PXRD patterns for $K_{0.8}Fe_{2-x}Co_xSe_2$ (nominal compositions x=0, 0.005, 0.015, 0.02, 0.025, 0.03, 0.035) were collected at room temperature. No impurities were detected within the resolution of x-ray diffractometer. All the peaks are well indexed to the tetragonal phase of $ThCr_2Si_2$-type (space group $I/4mmm$), indicating that all the samples are single phase. We adopted the structure model of $KFe_2Se_2$ to perform a Rietveld refinement [10]. The refined lattice constants $a$ and $c$ remain nearly unchanged considering the very low doping level of Co. The x-ray diffraction analysis of $K_{0.8}Fe_{1.98}Co_{0.02}Se_2$ crystal was shown in Fig. 1, which exhibits only 00$l$ ($l$=2n) diffraction peaks, indicating the cleavage surface is perfectly perpendicular to the *c*-axis. The SEM image of the crystal (the inset of Fig. 1) clearly shows the layered morphology.

Figure 2 displays the temperature dependence of in-plane electrical resistivity for $K_{0.8}Fe_{2-x}Co_xSe_2$ compounds with various concentrations of Co. The ρ at 60 K decreases gradually from 0.76 Ω·cm to 0.12 Ω·cm with the increase of Co doping, which may be owing to the increase of carriers brought by the Co substitution. Attention should be paid to the phenomenon that the normal state resistivity is much larger than other iron-based superconductors, which is probably due to the Fe vacancies in the conducting FeSe layers resulting in a strong carrier scattering [24]. The $T_c$ decreases monotonously with the increase of Co-substitution. It is suppressed very rapidly at the rate of $\Delta T_c$/Co-1 at. % (nominal) =-16 K, which is the fastest one ever-reported among the iron-based superconductors [20,21]. At the nominal concentration x≥0.03, no more zero resistivity is observed down to 5 K, although it shows an onset around T = 9 K. The inset shows the temperature-dependent resistivity curve for $K_{0.8}Fe_{1.975}Co_{0.025}Se_2$ at the temperature range from 5 K to 225 K, which stands for a general characteristic for all the superconducting samples. It is found that all of them have a semiconductor-to-metal like transition at about 150 K, which may correspond to the ordering of cation vacancies in this non-stoichiometric compound rather than structural or magnetic phase transition [25]. It should be noted that the resistance hump still exist after the disappearance of superconductivity, in contrast with the coexistence of superconductivity and the resistance hump in $K_{0.8}Fe_{1.7}Se_2$ under high pressure [14]. Currently, the exact correlation of the superconductivity and the resistance hump is not unambiguous. Further study is still needed.

Figure 3 shows the temperature dependence of susceptibility for the samples measured under a magnetic field of 40 Oe along the *c*-axis in zero-field-cooled (ZFC) and field-cooled (FC) processes. It can be observed that the diamagnetic response moves to lower temperatures as the increase of Co-doping. When the Co content is above 0.5 at. % (nominal 1.5 at. %), the superconducting transition disappears down to 5 K, which is in agreement with the results of the resistivity measurements. It is noticed that Meissner effect (FC) saturate at a much smaller value than shielding effect (ZFC), which is likely due to a strong pinning effect from the Co dopant in FeSe layers. Such rapid quenching of superconductivity reveals different mechanisms for the suppression of superconductivity in $K_{0.8}Fe_2Se_2$ and FeSe and further confirms that Co is indeed doped into the compounds.

In addition, doping dependence of the transition width ($\Delta T_c = T_c^{onset} - T_c^{zero}$) was shown in the inset of Fig. 4. They slightly broaden with the increasing concentration of Co. The superconducting transition temperatures are plotted as a function of nominal Co content (x) in Fig. 4. It can be seen that $T_c$ decreases quickly with a rate of about -16 K per nominal 1 at. % Co and the superconductivity is quenched by only 0.5 at. % measured Co's substitution, while the superconductivity in FeSe is suppressed at a relatively higher concentration of Co (5 at. %) [20,21]. NMR measurement in $(Fe_{0.9}Co_{0.1})Se$ showed that the strong spin fluctuations disappear and the superconductivity is strongly suppressed by the collapse of the Fermi surface nesting due to the electron doping introduced by the Co substitution [22]. However, for $K_{0.8}Fe_{2-x}Se_2$, the Fermi surface nesting is absent [26,27]. Therefore, the mechanism for the suppression of superconductivity in $K_{0.8}Fe_{2-x}Co_xSe_2$ is different from the case with FeSe.

For most iron pnictide superconductors, the 3$d$ electrons exhibit itinerant character [16,28]. In an itinerant model, within a simple rigid-band approach, Co doping is expected not to change the Fermi surface but just to shift the Fermi level, which means that only the total amount of electrons is relevant [16]. Recently, ARPES showed that the band around M in k-space is rather flat and shallow in $K_{0.8}Fe_2Se_2$, indicating a non-rigid band behavior [26]. Furthermore, the optical spectroscopy measurement on $K_{0.8}Fe_{2-x}Se_2$ demonstrated that the superconductivity is closely proximate to an antiferromagnetic semiconducting phase [29], whereas the parent compounds in other iron pnictides/ chalcogenides are spin-density-wave metals. Obviously, the itinerant model is not suitable to describe the rapid suppression of superconductivity in $K_{0.8}Fe_{2-x}Co_xSe_2$. In a scenario with localized 3$d$ electrons, the resulting behavior from the substitution of Fe by Co should be substantially different since the correlations in the Fe (Co) layers are directly influenced. Similarly, the substitution on the Cu site in the cuprates largely alters the band structure of the parent compound due to strong electron correlation effects, which leads to a fast suppression of superconductivity. Very recently, the parent compound of the (K,Tl)Fe$_x$Se$_2$ superconductors was proposed to be a Mott insulator at ambient pressure [30], similar to cuprates. From all the results mentioned above, we propose that the approach based upon localized 3$d$ model is more appropriate for $K_{0.8}Fe_2Se_2$. Certainly, the suppression may also be related to the destruction of the spin fluctuations relating to AFM, whereby copper pairs form. However, more efforts are still needed to clarify the issue.

In conclusion, we synthesized $K_{0.8}Fe_{2-x}Co_xSe_2$ (nominal x=0, 0.005, 0.015, 0.02, 0.025, 0.03, 0.035) by two step solid state reaction and characterized the resistivity and susceptibility of the freshly cleaved samples with various compositions. The superconductivity was found to be dramatically suppressed by rare substitution of Co. It implied that a localized model may be more appropriate in $K_{0.8}Fe_2Se_2$ superconductors. Such substitution of Fe by Co offers new clues to better understand the mechanism of superconductivity in the newly discovered $K_{0.8}Fe_2Se_2$ superconducting system.

This work was partly supported by the National Natural Science Foundation of China under Grants No. 90922037, 50872144 and 50972162 and the International Centre for Diffraction Data (ICDD, USA).

TABLE Ⅰ. ICP-AES analysis of elemental composition ratios for a series of $K_{0.8}Fe_{2-x}Co_xSe_2$.

FIG. 1. The x-ray diffraction pattern of $K_{0.8}Fe_{1.98}Co_{0.02}Se_2$ crystal indicates that the $00l$ ($l=2n$) diffraction peaks dominate the pattern. Inset shows the SEM image of the crystal.

FIG. 2 (color online). The temperature dependence of $ab$-plane resistance for $K_{0.8}Fe_{2-x}Co_xSe_2$ crystal. The inset is the $R_{ab}$-$T$ data for $K_{0.8}Fe_{1.975}Co_{0.025}Se_2$ from 5 K to 225 K.

FIG. 3 (color online). The magnetization of $K_{0.8}Fe_{2-x}Co_xSe_2$ crystal as a function of temperature with applied magnetic field parallel to $c$ axis.

FIG. 4 (color online). Variation of $T_c^{onset}$, $T_c^{mid}$ and $T_c^{zero}$ with Co content for $K_{0.8}Fe_{2-x}Co_xSe_2$. Inset shows the Co content dependence of the transition width.

| Nominal Co content(x) | Measured composition atomic ratio (K:Fe:Co:Se) |
| --- | --- |
| 0 | 0.39:0.85:0:1 |
| 0.005 | 0.39:0.83:0.001:1 |
| 0.015 | 0.41:0.86:0.002:1 |
| 0.02 | 0.4:0.83:0.003:1 |
| 0.025 | 0.39:0.84:0.005:1 |
| 0.03 | 0.39:0.84:0.005:1 |
| 0.035 | 0.39:0.83:0.006:1 |

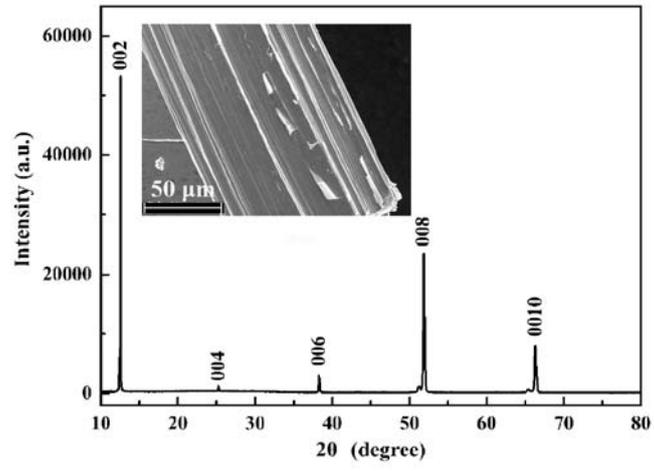

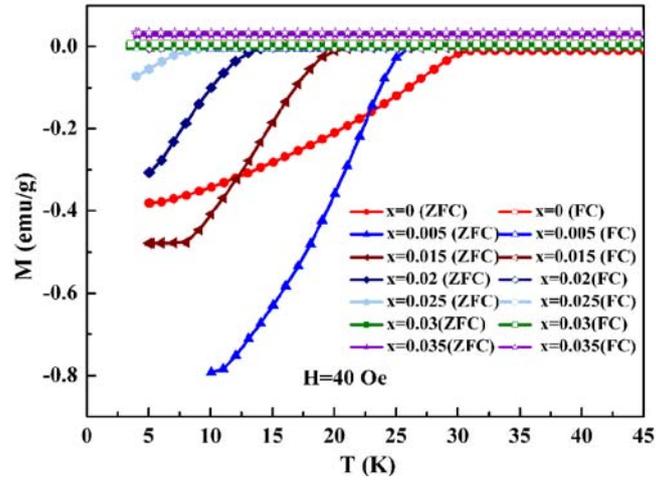

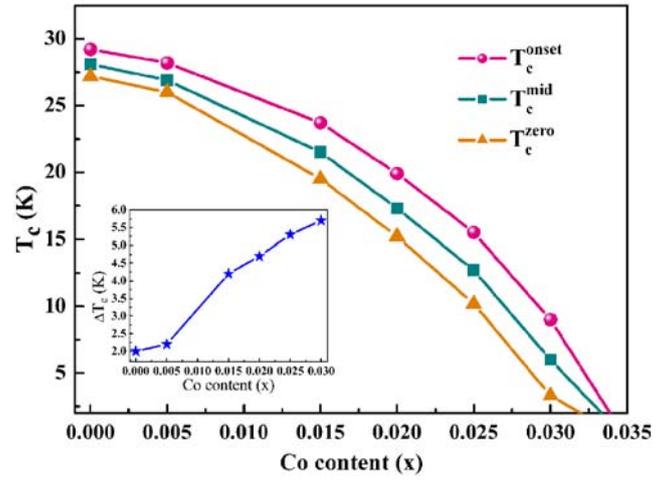